\documentclass[aps,prl,twocolumn,showpacs,superscriptaddress]{revtex4-1}  % for review and submission

\usepackage{upgreek}
\usepackage{graphicx,import}  % needed for figures
\usepackage{dcolumn}   % needed for some tables
\usepackage{bm}        % for math
\usepackage{color}
\usepackage{amssymb}  
\usepackage{amsmath}  
\usepackage{braket}

\newcommand{\updownarrows}{\mathbin\uparrow\hspace{-.3em}\downarrow}
\newcommand{\downuparrows}{\mathbin\downarrow\hspace{-.3em}\uparrow}
\renewcommand{\upuparrows}{\mathbin\uparrow\hspace{-.3em}\uparrow}
\renewcommand{\downdownarrows}{\mathbin\downarrow\hspace{-.3em}\downarrow}

\begin{document}
% The following information is for internal review, please remove them for submission
\widetext
%\leftline{V.3 To be submitted to PRL}
% the following line is for submission, including submission to the arXiv!!
%\hspace{5.2in} \mbox{Fermilab-Pub-04/xxx-E}

\title{Shuttling of Rydberg ions for fast entangling operations}
%\input author_list.tex       % D0 authors (remove the first 3 lines
                             % of this file prior to submission, they
                             % contain a time stamp for the authorlist)
                             % (includes institutions and visitors)
\author{J. Vogel}
\email{vogel@uni-mainz.de}  
\affiliation{QUANTUM, Johannes Gutenberg-Universit\"at Mainz, Staudinger Weg 7, 55128 Mainz, Germany}                   
\author{W. Li}
\affiliation{School of Physics and Astronomy and Centre for the Mathematics and Theoretical Physics of Quantum Non-equilibrium Systems, Nottingham, NG7 2RD, United Kingdom}    
\affiliation{Centre for the Mathematics and Theoretical Physics of Quantum Non-equilibrium Systems, Nottingham, NG7 2RD, United Kingdom}                 
\author{A. Mokhberi}      
\affiliation{QUANTUM, Johannes Gutenberg-Universit\"at Mainz, Staudinger Weg 7, 55128 Mainz, Germany}               
\author{I. Lesanovsky}
\affiliation{School of Physics and Astronomy and Centre for the Mathematics and Theoretical Physics of Quantum Non-equilibrium Systems, Nottingham, NG7 2RD, United Kingdom}  
\affiliation{Centre for the Mathematics and Theoretical Physics of Quantum Non-equilibrium Systems, Nottingham, NG7 2RD, United Kingdom}              
\author{F. Schmidt-Kaler}
\affiliation{QUANTUM, Johannes Gutenberg-Universit\"at Mainz, Staudinger Weg 7, 55128 Mainz, Germany}
\affiliation{Helmholtz-Institut Mainz, 55128 Mainz, Germany}
\date{\today}

\begin{abstract}
We introduce a scheme to entangle Rydberg ions in a linear ion crystal, using the high electric polarizability of the Rydberg electronic states in combination with mutual Coulomb coupling of ions that establishes common modes of motion. After laser-initialization of ions to a superposition of ground-  and Rydberg-state, the entanglement operation is driven purely by applying a voltage pulse that shuttles the ion crystal back and forth. This operation can achieve entanglement on a sub-$\upmu$s timescale, more than two orders of magnitude faster than typical gate operations driven by continuous-wave lasers. Our analysis shows that the fidelity achieved with this protocol can exceed $99.9\%$ with experimentally achievable parameters.
\end{abstract}

%\pacs{37.10.Pq, 37.10.Ty, 37.10.Mn}% Trapping of molecules, Ion trapping, Slowing and cooling of molecules.

\maketitle

In Rydberg states of an atom a valence electron is excited to a state with a high principal quantum number, leading to extraordinary large polarizabilities \citep{gallagher94a} and making them extremely susceptible to electric fields. Such high electric field susceptibility was employed for electric field sensing \cite{osterwalder99a,facon16a,penasa16a} and quantum information processing \cite{haroche06a}. For pairs, or arrays of atoms, a mutual electrical dipolar interaction of Rydberg states may lead to a blockade mechanism, which was proposed for generating  entanglement \cite{jaksch00a}. Pioneering experiments realized blockade driven entanglement with pairs of Rydberg atoms in optical tweezers \citep{isenhower10a,wilk10a}. Lately, arrays of Rydberg atoms \citep{gross17a} or atoms in reconfigurable optical tweezer potentials \cite{bernien17a,labuhn16a} have been used and allowed for remarkable progress in quantum simulation \citep{saffman10a}.    

More recently, trapped ions excited to Rydberg states \cite{feldker15a,higgins17a,higgins17b} have been investigated for exploring their unique features. The large electric polarizability has been characterized by spectroscopy and electric field mapping was exploited to position a single ion precisely inside the electric trap \cite{feldker15a,mokhberi19a}. Moreover, it has been shown that transitions to Rydberg states can be driven coherently from low-lying electronic states \cite{higgins19a}. A gate operation to entangle trapped Rydberg ions via a dipole-dipole interaction has been proposed, but requires microwave dressing of Rydberg states to cancel their polarizability \cite{mueller08a,li13a}.

\begin{figure}[]
\centering
\def\svgwidth{\columnwidth}
\footnotesize
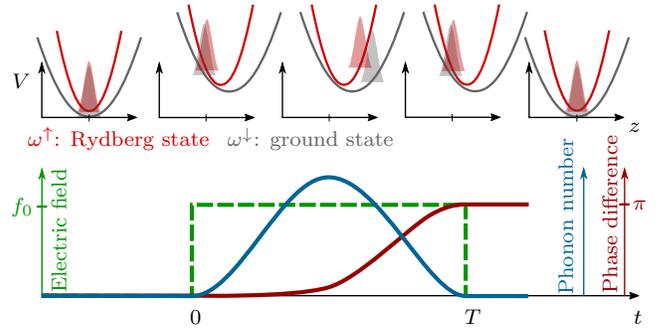
\caption{\textbf{Scheme for shuttle-based state-dependent phase accumulation.} Time evolution (from left to right) of the ion wavepacket in the presence of a fast electric kick with field-sensitive internal states. Confinement for ion in Rydberg state $\omega^\uparrow$ (red) is modified as compared to electronic ground state $\omega^\downarrow$ (gray). Ion displacement out of its equilibrium position by fast electric kick (green). Accumulated state-dependent phase difference (dark red) between Rydberg state and ground state, here $\pi$. Coherent motional excitation (blue) can be reduced to zero by adapting the pulse. }
\label{fig:EfieldGateSchematics}
\end{figure}

Here, we propose a scheme for entangling a pair of trapped ions, where we utilize unique features of this Coulomb crystals in Rydberg states: the electric polarizability and the corresponding energy shift of Rydberg states by an impulsive electric field. We design electric field waveforms that kick the two-ion crystal and impose a state-dependent force on common modes of motion. The shuttling of the crystal \cite{walther12a,bowler12a} leads to a  geometric phase, which can be controlled using the Rydberg principle quantum number $n$, the trap parameters and the shape of the kick. This entanglement operation is driven solely electrically and its duration may be as short as a few hundred ns, much faster than typical light-driven gates for ions \cite{kirchmair09a,benhelm08a,ballance16a,gaebler16a} and competing with gate operation times driven by pulsed laser sources \cite{wongcampos17a}.  It resembles laser-less ion entanglement operations driven by either static \cite{khromova12a} or dynamic magnetic gradients \cite{warring13a,weidt16a,harty16a} on the spin states of ions, however, driving large electric field gradients and performing strong electric kicks is an established technology in Paul traps. 

In the following, we sketch the state-dependent force for a single kicked ion and fully discuss the case of a two-ion crystal. We continue with the description of an entangling operation for a two-ion crystal. Furthermore, we describe the dominating sources of imperfections and optimize the shape of the electric kick. We conclude with a feasibility study, taking into account typical experimental parameters.        

\textbf{Spin-dependent electric kick.} We consider a single ion in a linear Paul trap, where a combination of radio-frequency and static electric fields generate a three-dimensional harmonic confinement. We are interested in the motion of the ion along the trap axis, the axis of weakest confinement, which is described by a harmonic oscillator with frequency $\omega$. Exciting an ion to a high-lying Rydberg state modifies the effective confinement due to the high polarizability \cite{higgins19a} - one may think of modifying its effective mass - such that the trap frequency $\omega^\alpha$ becomes state-dependent, where $\alpha=\{\uparrow, \downarrow\}$ denotes Rydberg state or ground state, respectively. Applying an electric kick displaces the ion out of its equilibrium position, introduces contributions from the induced electric dipole force but also drives the harmonic oscillator into vibrational excitation. A state-dependent phase is accumulated, see Fig.~\ref{fig:EfieldGateSchematics}. The coherently excited motion can be reduced to the initial state by properly choosing the pulse amplitude $f(t)$ and pulse duration $T$. A phase difference between Rydberg state and ground state is acquired. 

\textbf{Entanglement operation.} For two ions we control the phase of the electronic basis states $\ket{\alpha \beta}$ = $\{\ket{\downdownarrows}$, $\ket{\downuparrows}$, $\ket{\updownarrows}$,  $\ket{\upuparrows}\}$. The Coulomb interaction between the ions leads to state-dependent collective frequencies $\omega_j^{\alpha\beta}$ with the mode index $j=1,2$ where $\alpha\beta$ denotes the internal states of both individual ions, either in ground state or Rydberg state (see suppl. information): 
\begin{align}
\left(\omega_j^{\alpha\beta}\right)^2 &=\omega^{\alpha}\omega^{\beta}\left[\left(\frac{\omega^{\beta}}{\omega^{\alpha}}\right)^{\mspace{-10mu}(-1)^j} \mspace{-20mu}\cos^2 \theta^{\alpha\beta} +  \left(\frac{\omega^{\alpha}}{\omega^{\beta}}\right)^{\mspace{-10mu}(-1)^j} \mspace{-20mu}\sin^2 \theta^{\alpha\beta}\right]  \nonumber \\
						 &\quad+ (J^{\alpha\beta})^2 \left[1+ (-1)^j\sin(2\theta^{\alpha\beta}) \right] \\
(J^{\alpha\beta})^2 &= \frac{2(\omega^\alpha \omega^\beta )^2}{(\omega^\alpha )^2+(\omega^\beta )^2} \nonumber \\
\theta^{\alpha\beta} &= \frac{\pi}{4}-\frac{1}{2} \arctan \frac{(\omega^{\alpha} )^2-(\omega^{\beta} )^2}{2 (J^{\alpha\beta})^2} \nonumber 
\end{align}
Rydberg excitations in the ion crystal will affect the electric potential on neighboring ions, which leads to asymmetrical vibration around the center-of-mass due to a difference of effective masses \cite{morigi01a}. The potential energy is expressed in terms of the state-dependent creation $\tilde{a}_j^\dagger=(a_j^{\alpha\beta})^\dagger$ and annihilation  $\tilde{a}_j=a_j^{\alpha\beta}$ operators ($\hbar=1$).
\begin{align}
  H_\text{p}= \sum_{\alpha\beta=\uparrow,\downarrow}\left( \sum^2_{j=1} \omega_j^{\alpha\beta} \tilde{a}_j^\dagger \tilde{a}_j +V_0^{\alpha\beta}\right)\Pi^{\alpha\beta}
\end{align}
$V_0^{\alpha\beta}$ depends on the equilibrium positions of the ions, $\Pi^{\alpha\beta} =\ket{\alpha}_1\bra{\alpha}_1 \otimes \ket{\beta}_2\bra{\beta}_2$ is the projection operator.

Fast switching of an additional electric field $f(t)$ kicks the ions out of their equilibrium positions and drives the harmonic oscillator. The interaction of the electric field with the ion crystal can be described by a state-dependent kick $F_j^{\alpha\beta}(t)=f(t) l_j^{\alpha\beta} [\cos \theta^{\alpha\beta} - (-1)^j \sin \theta^{\alpha\beta}]$ acting on the vibrational mode with oscillator length $l_j^{\alpha\beta}=\sqrt{\hbar/(2m\omega_j^{\alpha\beta})}$ \citep{cirac00a,garcia-ripoll03a,garcia-ripoll05a}. Specifically, for ion crystals containing Rydberg excitations, we obtain the driving Hamiltonian
\begin{align}
	\mspace{-10mu} H_\text{d}(t)&=\mspace{-5mu} \sum_{\alpha\beta}\mspace{-5mu}\left[ \sum^2_{j=1} (F_j^{\alpha\beta}(t) \hspace{1mm} \tilde{a}_j + \text{h.c.}) + f(t)\;Z_\text{c}^{\alpha\beta} \right]\Pi^{\alpha\beta}. \label{eq:drivingHamiltonian}
\end{align}
The second term of Eq. (\ref{eq:drivingHamiltonian}) is proportional to the crystal center $Z_\text{c}^{\alpha\beta}$ and only affects the phase evolution of ion crystals with one Rydberg excitation, see suppl. information. The analytic time evolution operator $U_\text{I}$ for the driven harmonic oscillator is obtained using a Magnus expansion \citep{blanes10a}.
\begin{align}
	U_\text{I}(t) &= \sum_{\alpha\beta} \prod_{j=1}^2 \left[\mathcal{D}\left(A_j^{\alpha\beta}(t)\right)\right] \nonumber \\
			&\quad \times \exp \left[i \sum_{j=1}^2 \varphi_j^{\alpha\beta}(t) -i \Phi_\text{e}^{\alpha\beta}(t) \right] \Pi^{\alpha\beta} \label{eq:TimeEvolution}
\end{align}
The first term describes coherently generated vibrational mode excitation $A_j^{\alpha\beta}$ using the displacement operator $\mathcal{D}(A_j^{\alpha\beta})=\exp(A_j^{\alpha\beta}\tilde{a}_j + \text{h.c.})$. The total phase $\phi^{\alpha\beta}:=\varphi_1^{\alpha\beta}+\varphi_2^{\alpha\beta}+\Phi_\text{e}^{\alpha\beta}$ that is accumulated by each of the four basis states contains contributions from the vibrational modes and the crystal center displacement, respectively. 
\begin{figure}[t]
	\centering
	\def\svgwidth{\columnwidth}
	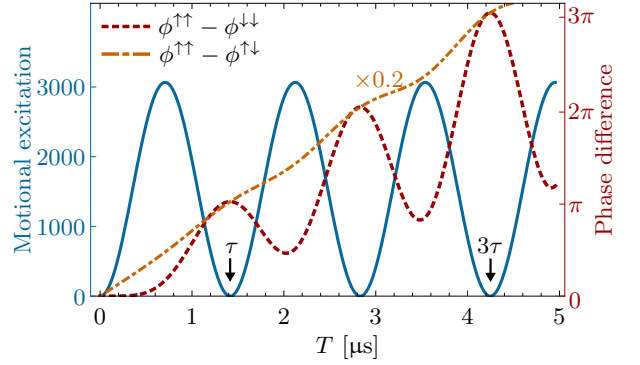	
	\caption{\textbf{Case study for controlled phase gate.} Coherent motional excitation measured as the number of phonons (blue) and relative phase accumulation (red) as a function of gate duration for Rydberg state $64$P in $^{40}$Ca$^+$. Kick shape is chosen to minimize residual motional excitation and generate phase differences of odd multiples of $\pi$ for state $\ket{\upuparrows}$ as compared to the states $\ket{\downdownarrows}$ (dark red, dashed) and $\ket{\updownarrows}$ (orange, dotdashed, scaled by $0.2$) at gate times indicated by arrows.}
	\label{fig:PhononAndPhase}
\end{figure}
Assuming a constant driving $f(t)=f_0$ for time $t\in[0,T]$, we obtain quantities from Eq. (\ref{eq:TimeEvolution}):
\begin{align}
	A_j^{\alpha\beta}(f_0,\omega_j^{\alpha\beta},T) &= f_0 \frac{l_j^{\alpha\beta}}{\omega_j^{\alpha\beta}} \left(e^{-i\omega_j^{\alpha\beta} T}-1 \right) \nonumber \\ 
	&\quad \times \left[\cos \theta^{\alpha\beta} - (-1)^j \sin \theta^{\alpha\beta}\right], \label{eq:PhononNumber}\\
	\varphi_j^{\alpha\beta}(f_0,\omega_j^{\alpha\beta},T) &= f_0^2  \left(\frac{l_j^{\alpha\beta}}{\omega_j^{\alpha\beta}} \right)^2 \left[\omega_j^{\alpha\beta} T -\sin \left( \omega_j^{\alpha\beta} T\right)\right] \nonumber \\ 
	&\quad \times \left[\cos \theta^{\alpha\beta} - (-1)^j \sin \theta^{\alpha\beta}\right]^2, \label{eq:PhaseTerm1}\\
	\Phi_\text{e}^{\alpha\beta}(f_0,\omega_j^{\alpha\beta},T) &= \left(f_0 Z_\text{c}^{\alpha\beta} + V_0^{\alpha\beta} \right) T.   \label{eq:PhaseTerm2}
\end{align}
The significance of the analytical equations (5-7) is that the entanglement operation is controlled only by the kick shape ($f_0,T$) and the common mode frequencies $\omega_j^{\alpha\beta}$. Therefore, arbitrary phase rotations and entanglement generation can be realized. For a controlled phase gate with two ions, we consider a phase difference $\phi^{\uparrow\uparrow}-\phi^{\downarrow\downarrow}=\pi$ while $\phi^{\downarrow\downarrow}=\phi^{\uparrow\downarrow}=\phi^{\downarrow\uparrow}$ and no residual excitation in phonon modes, thus $A_j^{\alpha\beta}=0$.

\textbf{Case study and experimental feasibility for $^{40}$Ca$^+$ ions.} In the case study, we consider Rydberg $n$P$_{1/2}$ states with a scalar polarizability $\mathcal{P}\propto n^7$. The state-dependent trap frequency is $\omega^\uparrow =\sqrt{\left(\omega^\downarrow \right)^2+\Delta\omega^2}$ with $\Delta\omega^2 =-16 \gamma^2\mathcal{P}/m$ \citep{li12a,li13a} and $\omega^\downarrow=2\sqrt{e\gamma/m}$, where $e$ is the electric unit charge and $\gamma$ the field gradient of the Paul trap. The trap frequencies are modified by the interaction between the highly excited electron with the core charge of the other ion and the large field sensitivity of Rydberg ions. Note that the relative frequency differences are $\leq 10^{-4}$, such that the excitation of the center-of-mass modes dominates, with a small excitation of the stretching mode for state $\ket{\uparrow\downarrow}$. For each vibrational mode, $ A_j^{\alpha\beta}$ is periodic and can be minimized by choosing a gate time $\tau=2\pi/\omega_j^{\alpha\beta}$, see Fig.~\ref{fig:PhononAndPhase}. Taking $n=64$ and $\omega_1^{\uparrow\uparrow}=2\pi \cdot 0.71\,$MHz we realize a controlled phase gate at $\tau$=1.4~$\upmu$s and $3\tau=$4.2~$\upmu$s (indicated by black arrows) with mitigated coherent excitation in the center-of-mass mode for state $\ket{\upuparrows}$ and correct relative phases. 

\begin{figure}[t]
	\centering
	\def\svgwidth{\columnwidth}
  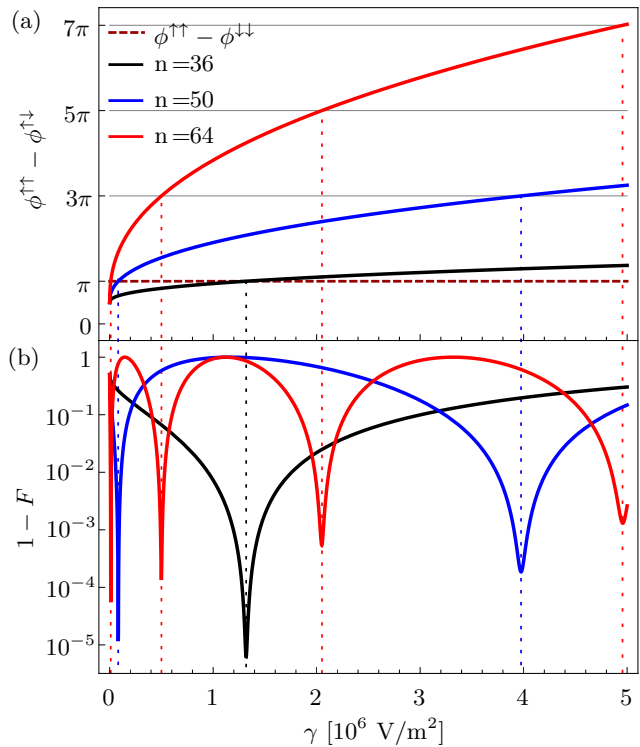
	\caption{\textbf{Differential phase and fidelity.} (a) Relative phase between states $\ket{\upuparrows}$ and $\ket{\updownarrows}$, and (b) infidelity dependent on the field gradient of the Paul trap for different principal quantum numbers of the Rydberg state. For every field gradient of the Paul trap the electric kick shape is chosen to realize $\phi^{\uparrow\uparrow}-\phi^{\downarrow\downarrow}=\pi$ (dashed dark red), and minimize residual phonons.}
	\label{fig:optimizedFidelity}
\end{figure} 

Varying the vibrational frequency by the field gradient of the Paul trap, the relative phase of $\pi$ between states $\ket{\upuparrows}$ and $\ket{\updownarrows}$ is modified, see Fig.~\ref{fig:optimizedFidelity}(a). With a specific combination of electric kick time and amplitude, we achieve a phase difference $\phi^{\uparrow\uparrow}-\phi^{\downarrow\downarrow}=\pi$, and minimize coherent excitation of modes. Rydberg states with higher principal quantum number and larger polarizability require smaller electric kick amplitude $f_0$ to acomplish the desired phase evolution. 

To characterize the entanglement operation, we analyze the state fidelity $F$ as the square of the overlap between a superposition $\ket{\Psi(0)}=1/2\left[(\ket{\downarrow}+\ket{\uparrow}) \otimes (\ket{\downarrow}+\ket{\uparrow})\right]$, initialized in the motional ground state and evolved under Eq.~(\ref{eq:TimeEvolution}), with the ideal target state. Fulfilling the phase conditions as indicated by the vertical dashed lines, the fidelity is limited by residual phonons due to the chosen electric kick. Optimizing the kick strength, kick duration and the ion confinement, a fidelity of $99.9\%$ can be achieved see Fig.~\ref{fig:optimizedFidelity}(b). For the 36P state the required electric kick strength is $E(t)=\frac{\hbar}{e}f(t)=28.75\,$V/m with a field gradient of the Paul trap of $\gamma = 1.32\cdot 10^6\,$V/m$^2$ and a vibrational mode frequency of $\omega_1^{\uparrow\uparrow}=2\pi \cdot 0.57\,$MHz, experimentally feasible with trapped Rydberg ions \cite{feldker15a,higgins17a}. Thereby, the ion crystal would be displaced along the trap axis by $10.9\,\upmu$m for a total operation time of $\tau$~=~1.76~$\upmu$s \cite{,walther12a,bowler12a}. The method comes with the advantage, that effects due to micromotion are mitigated, as the ion crystal moves along the trap axis. In principle, one might also employ a combination of axial and radial displacements for the gate, however, this will require synchronizing the electric kicking and oscillating radio-frequency field for the Paul trap as experimentally demonstrated in Ref.~ \cite{jacob16a}. The parasitic Stark shift from the applied electric field is estimated to be about four orders of magnitude smaller as compared to the separation of Rydberg energy levels, thus we do not anticipate state mixing and fidelity reduction.

\begin{figure}[t]
	\centering
	\def\svgwidth{\columnwidth}
	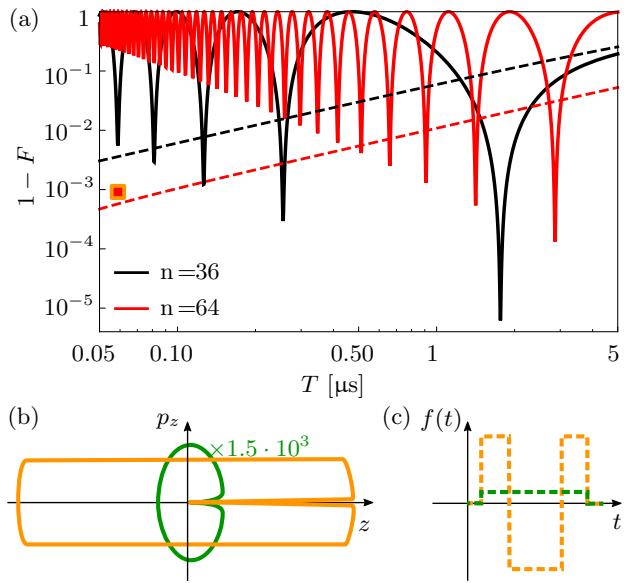
	\caption{\textbf{Optimized field kick.} (a) Entanglement fidelity and Rydberg state lifetime-limited fidelity (dashed) as a function of gate duration for $36$P (black) and $64$P (red) states of $^{40}$Ca$^+$. Here, we assume a lifetime of $65$~$\upmu$s and $370$~$\upmu$s, respectively. Bang-bang control: three consecutive kicks (red square with yellow contour) improve the fidelity to $99.9\%$ for $n=64$ at $60$ns operation speed. (b) Comparison of phase space trajectories and (c) field amplitudes for the single constant pulse (green, scaled by $1.5\cdot10^3$) and the waveform composed out of three kicks (yellow).}
	\label{fig:OptimalControl}
\end{figure}

\textbf{Lifetime limitation and optimized kicks.} A significant reduction of fidelity will arise from the finite Rydberg state lifetime, about 30$\,\upmu$s to 100$\,\upmu$s \cite{glukhov13a}, see dashed lines in Fig.~\ref{fig:OptimalControl}(a). Operation times above 1~$\upmu$s limit the fidelity to $90 \%-99\%$ depending on the Rydberg state. However, a bang-bang interaction by three consecutive kicks $f(t)=\{f_0,-f_0,f_0\}$ at times $t=\{0,T/4,3T/4\}$ leads to a fidelity of $99.9\%$ , see Fig.~\ref{fig:OptimalControl}(a-c). We  emphasize the importance of multi-kick sequences, in this example composed out of three kicks, as compared to the single constant pulse, see Fig.~\ref{fig:OptimalControl}(c). Note, that electric bang-bang control of single ions has been demonstrated with up to 10000 phonons and displacement pulses of sub-ns resolution \citep{alonso16a}, experimental parameters that even exceed the requirements for our proposed operation. Aiming for faster operations with higher fidelity, we will explore more complex phase trajectories of the wavepacket. The additional benefit of such schemes is a robustness against imperfections of the driving kick waveform. Such approaches have been discussed, however, in the context of the laser-driven M{\o}lmer-S{\o}rensen interaction \cite{schaefer18a,shapira18a} and might readily be adapted to our electric scheme. Alternatively optimal control theory may deliver optimized electric field waveforms \citep{caneva11a,fuerst14a,rach15a}.

\textbf{Conclusion and outlook.} In this work, we proposed a new scheme for fast entanglement operations based on electric kicks applied to trapped Rydberg ions in a linear Paul trap. Crucial to the successful implementation of this scheme is the high polarizability of $n$P Rydberg states that leads to a modification of the ion confinement and state-dependent vibrational modes. By tuning the field gradient of the Paul trap and shaping the electric kick, we optimize the scheme for entanglement operations of two ions. The parameter values required are well within regimes accessible by state-of-the-art experiments.

In future we may extend the scheme to linear ion crystals entangling more than two ions, or investigate spin-spin interactions in two-dimensional ion crystals \cite{nath15a} by state-dependent electric forces. In this context, we will study operations at finite temperature of the ion crystal, motional dephasing and heating by electric noise. The presented scheme may be adapted to the platform of neutral Rydberg atoms trapped in arrays of optical tweezers \cite{bernien17a,labuhn16a,keesling19a}. A set of common motions, analogous to the normal modes of vibration for the ion crystal, is established by the dipole-dipole interaction. State dependent forces between different Rydberg states can be implemented by a fast shuttling of the tweezer centers \cite{ohl19a} such that the trapped Rydberg atoms explore the AC-Stark shift from the tweezer potential, similar as the ion crystal explores the axial kick via its polarizability. In the array of Rydberg atoms the emerging collective energy-shifts may then be exploited to generate entanglement. We believe that experimental and theoretical work building on our ideas will be of relevance across a broad set of fields, such as multi-particle quantum systems with collective spin-motion coupling, quantum simulation and quantum information.

% Acknowledgement
\vspace{12pt}
We thank Rene Gerritsma for comments. This work was supported by the Deutsche Forschungsgemeinschaft (DFG) within the SPP 1929 Giant interactions in Rydberg Systems (GiRyd), the European Research Council under the European Union's Seventh Framework Programme (FP/2007-2013)  [ERC Grant Agreement No. 335266 (ESCQUMA)], within QuantERA by ERyQSenS, and by the EPSRC [Grant No. EP/R04340X/1]. A. M. acknowledges funding by Marie-Skłodowska-Curie grant agreement No. 796866 (Rydion).  I. L. gratefully acknowledges funding through the Royal Society Wolfson Research Merit Award. 

%For a single ion, the phonon excitation and phase accumulation is given by $|A_1^{\alpha\beta}|^2$ and $\varphi_1^{\alpha\beta}$ with $\theta=\pi/4$, evaluated at the trap frequency $\omega$ (see fig. 1). 

\bibliographystyle{apsrev4-1}
\bibliography{myreferencesMainz2}

\newpage
\appendix
\section{Supplementary information}
\subsection{Equilibrium positions and static potential energy}
The trap potential of two ions with state-dependent trap frequencies $\omega^\alpha$ with $\alpha=\{\uparrow,\downarrow\}$ resembling two possible internal states, reads
\begin{align}
V^{\alpha\beta}&=\frac{1}{2}m\left(\omega^\alpha\right)^2 Z_1^2 + \frac{1}{2}m\left(\omega^\beta\right)^2 Z_2^2 +\frac{\text{C}}{|Z_1-Z_2|} \nonumber \\
							 &\quad - 4e\gamma \left(Z_1 r_1 + Z_2 r_2 \right),
\end{align}
where $Z_j$ is the position of the $j$-th ion and $\text{C}=e^2/4\pi\epsilon_0$ is the Coulomb constant with $e$ the elementary charge, $\epsilon_0$ the vacuum permittivity, $\gamma$ the trap field gradient and $r_j$ the relative coordinate of the outer electron of each ion. We have introduced the interaction between the excited electron of a Rydberg state with the ion core by the last term. For small vibrations $z_j$ of the ionic core around the equilibrium position $\bar{Z}_j$, we expand the potential
\begin{align}
V^{\alpha\beta}&\approx z_1\left[m\left(\omega^\alpha\right)^2 \bar{Z}_1- \frac{\text{C}}{(\bar{Z}_1-\bar{Z}_2)^2} \right]\nonumber \\
&\quad+ z_2\left[m\left(\omega^\beta\right)^2 \bar{Z}_2+ \frac{\text{C}}{(\bar{Z}_1-\bar{Z}_2)^2}\right] \nonumber\\
&\quad+ V_\text{trap} + V_\text{DC}+ V_0^{\alpha\beta} +\mathcal{O}(z_1^n z_2^l), \nonumber\\
V_\text{trap}&= z_1^2\left[\frac{m\left(\omega^\alpha\right)^2}{2} + \frac{\text{C}}{|\bar{Z}_1-\bar{Z}_2|^3}\right] \nonumber \\
&\quad+ z_2^2\left[\frac{m\left(\omega^\beta\right)^2}{2} + \frac{\text{C}}{|\bar{Z}_1-\bar{Z}_2|^3}\right]-\frac{2\text{C}z_1z_2}{|\bar{Z}_1-\bar{Z}_2|^3}  \nonumber\\
V_\text{DC}&= z_1\left[-4e\gamma r_1 + \frac{2\text{C}(r_1 -r_2)}{|\bar{Z}_1-\bar{Z}_2|^3}\right] \nonumber \\
&\quad + z_2\left[-4e\gamma r_2 - \frac{2\text{C}(r_1 -r_2)}{|\bar{Z}_1-\bar{Z}_2|^3}\right]\nonumber 
\end{align}
where we kept up to quadratic expansions in terms of small quantities $r_j$ and $z_j$. $V_0^{\alpha\beta}$ gives a static potential, which depends on the equilibrium positions,
\begin{equation}
V_0^{\alpha\beta}=\frac{m\left(\omega^\alpha\right)^2 \bar{Z}_1^2}{2}+ \frac{m\left(\omega^\beta\right)^2 \bar{Z}_2^2}{2} +\frac{\text{C}}{|\bar{Z}_1-\bar{Z}_2|}
\end{equation}
As the change in equilibrium positions by the additional static potential $V_\text{DC}$ is typically small, we calculate the ion distances between two ground state ions:
\begin{align}
\frac{\text{C}}{|\bar{Z}_1-\bar{Z}_2|^3}=2e\gamma \nonumber 
\end{align}
and obtain:
\begin{align}
V_\text{DC} \approx -4e\gamma(r_2z_1+r_1z_2)
\end{align}
We identify this term as a cross-coupling between the negatively singly charged electron of one ion with the positively double charged core of the other ion. For two ions in the Rydberg state both ions see an additional trap potential. If only one ion is in the Rydberg state, the neighboring ground state ion will see a modified trap potential. The term $V_\text{DC}$ describes the electron dynamics, which is typically much faster than the ion core dynamics, and can be treated via second order perturbation. Thereby, we obtain modified trap frequencies for Rydberg ions compared to ground state ions (see main text).
From the force balance condition (linear orders) the state-dependent equilibrium positions are obtained
\begin{eqnarray}
\bar{Z}_1^{\alpha\beta}&&=\phantom{-}\left[\frac{\text{C}\left(\omega^\beta \right)^4}{m\left(\omega^\alpha \right)^2 \left(\left(\omega^\alpha \right)^2 +\left(\omega^\beta \right)^2 \right)}\right]^{\frac{1}{3}},\\
\bar{Z}_2^{\alpha\beta}&&=-\left[\frac{\text{C}\left(\omega^\alpha \right)^4}{m\left(\omega^\beta \right)^2 \left(\left(\omega^\alpha \right)^2 +\left(\omega^\beta \right)^2 \right)}\right]^{\frac{1}{3}},
\end{eqnarray} 
and we find the geometric center of the two ions: 
\begin{eqnarray}
Z_\text{c}^{\alpha\beta} &&= \bar{Z}_1^{\alpha\beta}+\bar{Z}_2^{\alpha\beta} \nonumber \\
&&= \frac{\text{C}^{1/3} \left(\left(\omega^\beta \right)^2-\left(\omega^\alpha \right)^2\right)}{\left[m\left(\omega^\beta \right)^2 \left(\omega^\alpha \right)^2 \left(\left(\omega^\alpha \right)^2+\left(\omega^\beta \right)^2\right)^2\right]^{1/3}}.
\end{eqnarray}
With the equilibrium positions, we can calculate the static potential
\begin{eqnarray}
V_0^{\alpha\beta}= \frac{3}{2}\left[\frac{m\text{C}^2\left(\omega^\alpha \right)^2 \left(\omega^\beta \right)^2}{\left(\omega^\alpha \right)^2+\left(\omega^\beta \right)^2}\right]^{\frac{1}{3}}
\end{eqnarray}
 
We should note that the state-dependent difference of the static potential $V_0^{\alpha\beta}$ will be canceled by proper laser detuning. This means that the laser frequency for excitation of a single ion and two ions to the Rydberg state will be different.

\subsection{Phonon modes}
Using the equilibrium positions, we obtain
\begin{eqnarray}
V^{\alpha\beta}&&=\frac{m z_1^2}{2}\left[\left(\omega^\alpha \right)^2 + \left(J^{\alpha\beta}\right)^2\right] \nonumber \\
&&\quad+\frac{m z_2^2}{2}\left[\left(\omega^\beta \right)^2 + \left(J^{\alpha\beta}\right)^2\right]\nonumber \\
&&\quad- m \left(J^{\alpha\beta}\right)^2 z_1z_2 +V_0^{\alpha\beta}, 
\end{eqnarray}
with 
\begin{equation}
\left(J^{\alpha\beta}\right)^2 = \frac{2\left(\omega^\alpha \omega^\beta \right)^2}{\left(\omega^\alpha \right)^2+\left(\omega^\beta \right)^2}
\end{equation}
We introduce a transformation, which mixes position coordinates by an angle $\theta$: 
\begin{equation}
\begin{pmatrix}
z_1 \\ z_2
\end{pmatrix} = 
\begin{pmatrix}
\phantom{-}\cos {\theta} & -\sin{\theta} \\ \sin{\theta} & \cos{\theta}
\end{pmatrix} \begin{pmatrix}
q_1 \\ q_2
\end{pmatrix},
\end{equation}
and the potential becomes
\begin{align}
V&=\frac{m}{2}\left(\left(\omega_1^{\alpha\beta} \right)^2q_1^2+\left(\omega_2^{\alpha\beta} \right)^2q_2^2 \right) +V_0^{\alpha\beta} \\
&\quad+ \frac{m}{2}\left[\left(\left(\omega^{\alpha} \right)^2-\left(\omega^{\beta} \right)^2\right)\sin2\theta -2\left(J^{\alpha\beta}\right)^2 \cos 2\theta \right]q_1q_2 \nonumber 
\end{align}
with
\begin{align}
\left(\omega_j^{\alpha\beta}\right)^2&=\omega^{\alpha}\omega^{\beta}\left[\left(\frac{\omega^{\beta}}{\omega^{\alpha}}\right)^{\mspace{-10mu}(-1)^j} \mspace{-20mu}\cos^2 \theta^{\alpha\beta} +  \left(\frac{\omega^{\alpha}}{\omega^{\beta}}\right)^{\mspace{-10mu}(-1)^j} \mspace{-20mu}\sin^2 \theta^{\alpha\beta}\right]  \nonumber \\
						 &\quad+ (J^{\alpha\beta})^2 \left[1+(-1)^j \sin(2\theta^{\alpha\beta}) \right]. 
\end{align}
The potential is diagonal when the mixing angle becomes
\begin{equation}
\theta^{\alpha\beta} = \frac{\pi}{4}-\frac{1}{2}\arctan\frac{\left(\omega^{\alpha} \right)^2-\left(\omega^{\beta} \right)^2}{2\left(J^{\alpha\beta}\right)^2}.
\end{equation}
At low temperatures, the vibrations are described by vibrational quanta acting on the collective coordinates $q_j=l_j^{\alpha\beta}\left(\tilde{a}_j^\dagger  + \tilde{a}_j \right)$ with the oscillator length $l_j^{\alpha\beta}=\sqrt{\hbar/\left(2m\omega_j^{\alpha\beta}\right)}$ and the phonon Hamiltonian is expressed in terms of the state-depending creation $\tilde{a}_j^\dagger=(a_j^{\alpha\beta})^\dagger$ and annihilation  $\tilde{a}_j=a_j^{\alpha\beta}$ operators ($\hbar=1$).
\begin{align}
H^{\alpha\beta}&=\sum^2_{j=1} \omega_j^{\alpha\beta} \tilde{a}_j^\dagger \tilde{a}_j +V_0^{\alpha\beta}.
\end{align}
Summing up all basis states we obtain the phonon Hamiltonian of the full system:
\begin{equation}
  H_\text{p}=\sum_{\alpha\beta=\uparrow,\downarrow}\left( H^{\alpha\beta}\right)\Pi^{\alpha\beta} 
\end{equation}
with the state projection operator
\begin{equation}
  \Pi^{\alpha\beta} =\ket{\alpha}_1\bra{\alpha}_1 \otimes \ket{\beta}_2\bra{\beta}_2
\end{equation}
\subsection{Electric kick}
The Hamiltonian for the fast electric pulse driving the ions is given by
\begin{eqnarray}
H_\text{d}&=&f(t)(Z_1+Z_2) \\
	&=& f(t) (z_1+z_2) + f(t)(\bar{Z}_1^{\alpha\beta}+\bar{Z}_2^{\alpha\beta}) \\
	&=& f(t)(\cos\theta^{\alpha\beta} +\sin\theta^{\alpha\beta})q_1 \nonumber \\
	& &\quad + f(t)(-\sin\theta^{\alpha\beta} +\cos\theta^{\alpha\beta})q_2 + f(t)Z_\text{c} \\
	&=& f(t)l_1^{\alpha\beta}(\cos\theta^{\alpha\beta} +\sin\theta^{\alpha\beta})\left(\tilde{a}_1^\dagger+\tilde{a}_1\right) \nonumber \\
	& &\quad + f(t)l_2^{\alpha\beta}(\cos\theta^{\alpha\beta} -\sin\theta^{\alpha\beta})\left(\tilde{a}_2^\dagger+\tilde{a}_2\right) \nonumber \\
	& &\quad + f(t)Z_\text{c}^{\alpha\beta} \\
	&=& \sum^2_{j=1} F_j^{\alpha\beta}(t)\left(\tilde{a}_j^\dagger+\tilde{a}_j\right)  + f(t)Z_\text{c}^{\alpha\beta}.
\end{eqnarray}
The driving Hamiltonian is given by:
\begin{align}
	\mspace{-20mu} H_\text{d}(t)&=\mspace{-5mu} \sum_{\alpha\beta}\mspace{-5mu}\left[ \sum^2_{j=1} (F_j^{\alpha\beta}(t) \hspace{1mm} \tilde{a}_j + \text{h.c.}) + f(t)\;Z_\text{c}^{\alpha\beta} \right]\Pi^{\alpha\beta}. 
\end{align}
\subsection{Time evolution operator}
We obtain the interaction Hamiltonian by:
\begin{align}
	H_\text{I} &= e^{i H_\text{p} t}\,H_\text{d}\, e^{-i H_\text{p} t}\nonumber \\
	H_\text{I}&=\mspace{-5mu}\sum_{\alpha\beta}\mspace{-5mu}\left( \sum^2_{j=1} \left(F_j^{\alpha\beta}(t)e^{i\omega_j^{\alpha\beta} t }\tilde{a}_j + \text{h.c.}\right)  \right. \nonumber \\ 
	&\quad+ \left. \vphantom{\sum^2_{j=1}} f(t)Z_\text{c}^{\alpha\beta} + V_0^{\alpha\beta} \right)\Pi^{\alpha\beta}.
\end{align}
As the Hamiltonian in the interaction picture is time dependent we use a Magnus expansion for time ordered systems:
\begin{align}
U_\text{I} (t) &= \exp \left[-i \int_{t_0}^t d\tau H_\text{I} (\tau) \right. \nonumber \\
			 &\quad- \frac{1}{2}\int_{t_0}^t d\tau ' \int_{t_0}^{\tau '} d\tau \left[H_\text{I} (\tau '),H_\text{I} (\tau)\right] \nonumber \\
			 &\quad+ \frac{i}{6}\int_{t_0}^t d\tau '' \int_{t_0}^{\tau ''} d\tau'  \int_{t_0}^{\tau '} d\tau \nonumber \\
			 &\quad\times \left( \left[H_\text{I} (\tau ''),\left[H_\text{I} (\tau '),H_\text{I} (\tau)\right]\right] \right.\nonumber \\
			 &\quad+ \left. \left. \left[H_\text{I} (\tau),\left[H_\text{I} (\tau '),H_\text{I} (\tau '')\right]\right]\right) +\cdots \vphantom{\int_{t_0}^t}\right] 
\end{align}
Using the commutator relations $\left[\tilde{a}_i,\tilde{a}_j^\dagger\right]=\delta_{ij}$ and $\left[\tilde{a}_i,\tilde{a}_j\right]=\left[\tilde{a}_i^\dagger,\tilde{a}_j^\dagger\right]=0$ of creation and annihilation operators we find 
\begin{align}
  &\left[H_\text{I} (\tau '),H_\text{I} (\tau)\right] = \sum_{\alpha\beta} \\
	&\left( - 2i \sum_{j=1}^2 F_j^{\alpha\beta}(\tau ')F_j^{\alpha\beta}(\tau) \sin \left[ \omega_j^{\alpha\beta}(\tau '- \tau)\right]\nonumber\right)\Pi^{\alpha\beta}  \\
	&\left[H_\text{I} (\tau ''),\left[H_{I} (\tau '),H_{I} (\tau)\right]\right] = 0
\end{align}
The complete time evolution operator is thereby:
\begin{align}
	U_\text{I}(t) &= \sum_{\alpha\beta} \prod_{j=1}^2 \left[\mathcal{D}\left(A_j^{\alpha\beta}(t)\right)\right] \nonumber \\
			&\quad \times \exp \left[i \sum_{j=1}^2 \varphi_j^{\alpha\beta}(t) -i \Phi_\text{e}^{\alpha\beta}(t) \right] \Pi^{\alpha\beta}
\end{align}
with
\begin{align}
A_j^{\alpha\beta}\left(f(t),\omega_j^{\alpha\beta},t\right) &= \int_{t_0}^{t} d\tau F_j^{\alpha\beta}(\tau) e^{i\omega_j^{\alpha\beta} \tau} ,\\
\Phi_\text{e}^{\alpha\beta}\left(f(t),\omega_j^{\alpha\beta},t\right)&=\int_{t_0}^{t} d\tau f(t)Z_\text{c}^{\alpha\beta} + V_0^{\alpha\beta}, \\
\varphi_j^{\alpha\beta}\left(f(t),\omega_j^{\alpha\beta},t\right)&=\int_{t_0}^t d\tau ' \int_{t_0}^{\tau '} d\tau \left[ F_j^{\alpha\beta}(\tau ')F_j^{\alpha\beta}(\tau) \right. \nonumber \\
& \quad\left. \times \sin \left( \omega_j^{\alpha\beta} (\tau ' - \tau)\right)\right].
\end{align}
\subsection{Gate fidelity}
We analyze the time evolution of the electronic basis states $\ket{\alpha \beta}$ = $\{\ket{\downdownarrows}$, $\ket{\downuparrows}$, $\ket{\updownarrows}$,  $\ket{\upuparrows}\}$. As the states $\ket{\downuparrows}$ and $\ket{\updownarrows}$ are symmetric, we only consider state $\ket{\updownarrows}$. 
An ideal controlled phase gate has the evolution operator
\begin{align}
	U_{\text{CP}}=\left(\begin{array}{rrrr}
		1 & 0 & 0 & 0 \\
		0 &\phantom{-}1 & 0 & 0 \\
		0 & 0 & \phantom{-}1 & 0 \\
		0 & 0 & 0 & -1 
		\end{array} \right) . \label{eq:CPgateIdeal}
\end{align}
The state fidelity without lifetime limit is defined as the overlap $F =|\bra{\Psi (0)} U_{\text{CP}} U_I(T) \ket{\Psi (0)}|^2$, which we evaluate here explicitly for the two ion superposition state $\ket{\Psi(0)}=1/2\left[(\ket{\downarrow}+\ket{\uparrow}) \otimes (\ket{\downarrow}+\ket{\uparrow})\right]$ with the ion crystal initially in the motional ground state. For a constant driving field we obtain:

\begin{eqnarray}
F&=&\frac{1}{16}\left|\exp\left[-\frac{\left|A_1^{\downarrow\downarrow}\right|^2}{2}+i\left(\varphi_1^{\downarrow\downarrow}+\Phi_\text{e}^{\downarrow\downarrow} \right)\right] \right. \nonumber \\
& &\quad + 2\exp\left[-\frac{\left|A_1^{\uparrow\downarrow}\right|^2}{2}-\frac{\left|A_2^{\uparrow\downarrow}\right|^2}{2}+i\left(\varphi_1^{\uparrow\downarrow}+\varphi_2^{\uparrow\downarrow}+\Phi_\text{e}^{\uparrow\downarrow} \right)\right]\nonumber \\
& &\quad - \left. \exp\left[-\frac{\left|A_1^{\uparrow\uparrow}\right|^2}{2}+i\left(\varphi_1^{\uparrow\uparrow}+\Phi_\text{e}^{\uparrow\uparrow} \right)\right]\right|^2.
\end{eqnarray}

\end{document}